\documentclass{iaus}
\usepackage{graphicx}

\title[AVOCADO] 
{AVOCADO: A Virtual Observatory Census to Address Dwarfs Origins}

\author[Rub\'en S\'anchez-Janssen]   
{Rub\'en S\'anchez-Janssen \and the AVOCADO collaboration.}

\affiliation{ESO, Alonso de C\'ordova 3107, Santiago, Chile}

\pubyear{2011}
\volume{277}  
\pagerange{119--126}
\jname{Tracing the Ancestry of Galaxies on the Land of our Ancestors}
\editors{Carignan, C., Freeman, K. C. \& Combes, F., eds.}
\begin{document}

\maketitle

\begin{abstract}
Dwarf galaxies are by far the most abundant of all galaxy types, yet their properties are still poorly understood --especially due to the observational challenge that their intrinsic faintness represents. AVOCADO aims at establishing firm conclusions on their formation and evolution by constructing a homogeneous, multiwavelength dataset for a statistically significant sample of several thousand nearby dwarfs ($-18 < M_{i} < -14$). Using public data and Virtual Observatory tools, we have built GALEX+SDSS+2MASS spectral energy distributions that are fitted by a library of single stellar population models. Star formation rates, stellar masses, ages and metallicities are further complemented with structural parameters that can be used to classify them morphologically. This unique dataset, coupled with a detailed characterization of each dwarf's environment, allows for a fully comprehensive investigation of their origins and to track the (potential) evolutionary paths between the different dwarf types.
\end{abstract}

\firstsection 
\section{Introduction}
Dwarf galaxies are the most abundant of all galaxy types, and this simple fact makes them already key objects to constrain galaxy formation and evolution models. It is currently well established that the low mass haloes where they reside are extremely inefficient in converting baryons to stars, as indicated by their observed high gas mass fractions (e.g., \cite{Geha+06}). However, despite this inefficiency, the peak of specific star formation rate (SSFR) shifts towards lower masses at lower redshifts (e.g., \cite{Martin+07}) --that is, the main epoch of star formation activity for low mass systems is \emph{now}, as opposed to more massive galaxies (the well known downsizing effect). Their high SSFR and low metallicities, together with the lack of spiral structure further make dwarf galaxies excellent laboratories for the study of star formation and feedback effects, with conditions likely closely resembling those at high redshift. Finally, dwarfs, having low masses and densities, are very sensible tracers of interactions and environmental processes, thus helping to understand the role that external mechanisms play in galaxy evolution.

Despite their critical importance, dwarf galaxy properties are still poorly understood --both from the theoretical and the observational side. Simulations cannot yet fully reproduce the process of dwarf formation, as they require challenging large resolutions to accurately describe the complex baryon physics that dominate these small scales. Thus, star formation occurs too early and fast in semi-analytic models of  dwarf formation, resulting in an excess of faint red systems with respect to observations (e.g., \cite{Henriques+08}). Moreover, all detailed hydrodynamical simulations produce dwarfs with too high baryonic masses for their corresponding dark matter haloes (\cite{Sawala+10}).
From an observational point of view, their intrinsic faintness prevents studies of statistically significant field dwarf samples, with the most complete ones barely exceeding the hundredth (e.g., \cite{HE06}).
This effect even worsens at higher redshifts ($z > 0.1$), where field dwarf galaxies are rather poorly characterized --although the knowledge is deeper in the case of cluster dwarfs, both for low (\cite{SJ+08}) and intermediate redshift samples (\cite{Barazza+09}).

AVOCADO (A Virtual Observatory Census to Address Dwarfs Origins) aims at providing strong constraints on dwarf formation and evolution by constructing a homogeneous, multiwavelength dataset for a statistically significant sample of several thousand nearby dwarfs. Here we present the dataset and the immediate goals of the project.

\section{Sample and Data}
Our parent sample is drawn from the NYU-VAGC DR7 sample (\cite{Blanton+05}), which has been further cross-correlated with NED to obtain additional redshifts when possible. We have selected as dwarfs all the objects having $2000 < cz <6000$ km\,s$^{-1}$ and $M_{i}-5\,log\,h_{100} > -18$. The lower recession velocity limit was set to avoid strong corrections from the Virgo infall velocity field, while the upper ensures that the sample is volume-limited for objects $M_{i}<-16$ and $<$$\mu$$>$$_{r,50}< 24$. While the $i$-band limit is somewhat arbitrary, it roughly corresponds to the scale where the break of the baryonic Tully-Fisher occurs (\cite{McGaugh+00}), and where galaxies start to be systematically thicker (\cite{SJ+10}), deviating from usual disk scaling relations (\cite{Schombert06}) --most likely a result of the increasing importance of turbulent motions due to feedback effects. The final sample consists of $\sim$7000 dwarf galaxies, with $\sim$1500 of them being fainter than $M_{i} = -16$.

\begin{figure}[t]
\begin{center}
\includegraphics[width=.9\textwidth]{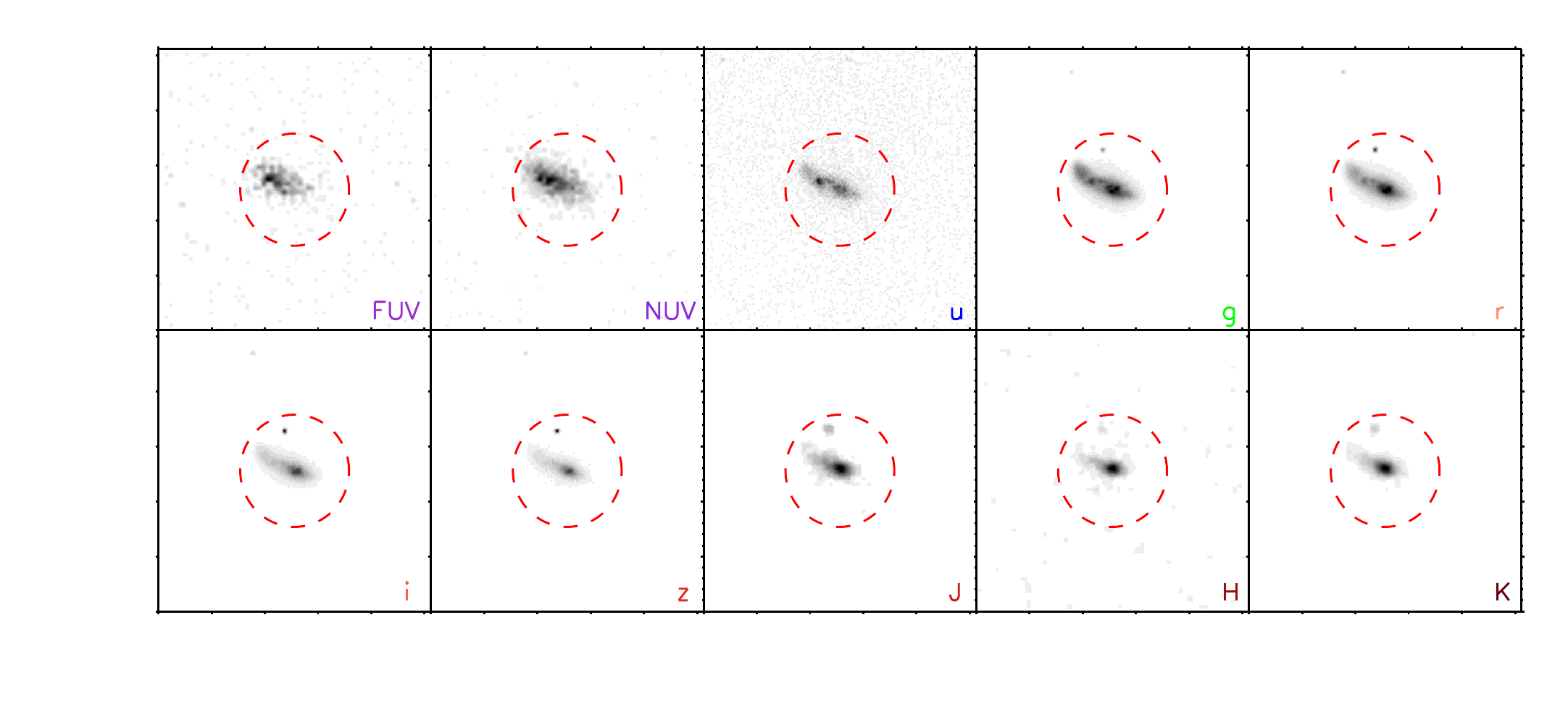} 
\caption{Example of postage-stamp images for a dwarf in the ten UV-to-NIR filters used in AVOCADO. The dashed circle indicates 2\,r$_{petro}$, the common aperture used to compute all fluxes.} 
\label{fig:stamps}
\end{center}
\end{figure}

For these objects we have retrieved GALEX+SDSS+2MASS  postage-stamp images centred on the object (see Fig.\,\ref{fig:stamps}). These images are used to compute Petrosian magnitudes within a common aperture for all filters, further allowing for measurements below the GALEX and 2MASS catalogs limits. 

\section{VOSA: the VO SED Analyzer}
Our ten-point UV-to-NIR spectral energy distributions (SEDs) are fitted to a library of single stellar populations (SSPs) using VOSA\,\footnote[1]{http://svo.cab.inta-csic.es/theory/vosa3/}, a Virtual Observatory (VO) tool of public use whose specific capabilities have been developed for this project. VOSA can be run with user-provided photometric catalogues, or public catalogues accessible from the VO can be queried. The tool includes two options to estimate the optimal model(s) reproducing the data: a traditional $\chi^{2}$ minimization, and a Bayesian approach (in the line of Kaufmann et al. 2003). While any VO-compliant stellar population model can in principle be used in the fitting procedure, AVOCADO uses the PopStar evolutionary synthesis models of \cite{Molla+09}. They feature a variety of IMFs, ages ($5 < \mbox{log(t/yr)} < 10.18$) and metallicities ($1/200 < \mbox{Z/Z$_{\odot}$} < 2.5$) and, most importantly, (optionally) include the emission of H+He nebular continuum --which significantly reddens colours of young, metal-poor SSPs, thus affecting their derived parameters (see, e.g., C. Maraston's contribution in these proceedings).

\section{Structure and Morphology}
Ideally, one would like to compare the properties derived from SED fitting for all different types of known dwarfs (dIrrs, dEs...). However, traditional visual classification is a slow and subjective (i.e., non-reproducible) procedure. If galaxies can be divided into different types, it is only because they occupy (more or less) delimited regions in a given parameter space (e.g., \cite{GdP+03}). Therefore, every galaxy in AVOCADO is characterized by a set of quantitative morphological parameters --including asymmetry, clumpiness, concentration, M$_{20}$, Gini, $b/a$ and $<$$\mu_{e}$$>$, among others. Colour is not used as a discriminator in order to decouple the structural information from the galaxy star formation activity. To ease the comparison between different types and with previous results, we again follow a statistical approach. We use the GALSVM code (\cite{HC+08}) to identify non-linear boundaries in the previous $n$-dimensional parameter space. The boundaries are defined by a set of dwarf training samples selected from the literature, broadly divided into four groups (dEs, dIrrs, BCDs and dSps). We then assign a probability that a given galaxy belongs to each group according to its location in the parameter space, so for every object $P(dE)+P(dIrr)+P(BCD)+P(dSp) = 1$. $P(Type)$ is therefore an indicator of how much a given dwarf resembles the typical galaxy of this $Type$.

\section{Analysis of SDSS spectra}
The analysis of SEDs will provide physical information for the whole sample of dwarfs, allowing for a detailed study of close companions and an optimal comparison with future, high-redshift studies that most probably will lack spectroscopic information. However, SDSS spectra are available for most of the AVOCADO sample, and we are analysing them with Starlight (\cite{CF+05}).  This public code decomposes each spectrum in a linear combination of several SSPs ($N<5$ in the case of AVOCADO), thus providing detailed constraints on the galaxy's stellar mass, dust extinction, ages, metallicities and SFRs. These parameters will be compared with those derived from SED fitting, so we can address the limitations imposed by photometric studies. Additionally, the continuum-subtracted spectra will allow the study of emission line equivalent widths (EWs) and line ratios to characterise the properties of the ionized gas in dwarfs.

\section{Environment}
The role that stellar mass and halo mass (environment) play in galaxy evolution is not yet fully understood. What is clear though is that central and satellite galaxies have strikingly different properties, suggesting that environmental effects are indeed relevant. Unfortunately, it is well known that environmental studies with the SDSS suffer from severe limitations for nearby galaxies, such as edge effects, bright galaxy shredding and spectroscopic bias against close companions. Therefore, in AVOCADO we have adopted the following approach.  We use the HyperLeda database to construct an all-sky sample of galaxies essentially complete down to $M_{B} = -19.5$, which is used to look for companions at scales $<$1 Mpc (see \cite{Blanton+06}). Once the companion producing the highest effect on the target galaxy is identified (by means of a tidal parameter, see \cite{SJ+10}), we use the SDSS to look for closer, fainter galaxies that could produce comparable effects. We therefore will have an unbiased estimation of both the density field on Mpc scales, as well as of the contribution of faint, close companions.

\section{Neutral and molecular gas}
The final ingredient still missing to draw a robust evolutionary scenario for dwarf galaxies is their neutral and molecular gas content.
For AVOCADO we have compiled H\,{\sc i} data from the literature for a few hundred dwarfs. A major breakthrough, however, will have to wait for the completion of already planned all-sky surveys, such as WALLABY and WNSHS (see G. Jozsa's contribution in these proceedings).
Furthermore, large CO surveys of AVOCADO dwarfs will be possible with the forthcoming advent of ALMA, with profound implications specially for faint, metal-poor dwarfs \cite{Leroy+05}.\\

The AVOCADO dataset is thus expected to become a benchmark for comparisons with numerical simulations and high redshift studies of dwarf galaxies.


\end{document}